\documentclass[apj]{emulateapj}

\def\gtorder{\mathrel{\raise.3ex\hbox{$>$}\mkern-14mu
             \lower0.6ex\hbox{$\sim$}}}
\def\ltorder{\mathrel{\raise.3ex\hbox{$<$}\mkern-14mu
             \lower0.6ex\hbox{$\sim$}}}

\usepackage{colordvi}
\usepackage{amsmath}
\usepackage{hyperref}

\usepackage{academicons}

\slugcomment{Accepted to PASP}

\shorttitle{LAST pipeline I}

\shortauthors{Ofek et al.}

\begin{document}

\title{The Large Array Survey Telescope -- Pipeline. I. Basic image reduction and visit coaddition}
\author{
E.~O.~Ofek\altaffilmark{1},
Y.~Shvartzvald\altaffilmark{1},
A.~Sharon\altaffilmark{1},
C.~Tishler\altaffilmark{1},
D.~Elhanati\altaffilmark{1},
N.~Segev\altaffilmark{1},
S.~Ben-Ami\altaffilmark{1},
G.~Nir,\altaffilmark{2},
E.~Segre\altaffilmark{1},
Y.~Sofer-Rimalt\altaffilmark{1},
A.~Blumenzweig\altaffilmark{1},
N.~L.~Strotjohann\altaffilmark{1},
D.~Polishook\altaffilmark{1},
A.~Krassilchtchikov\altaffilmark{1},
A.~Zenin\altaffilmark{3},
V.~Fallah~Ramazani\altaffilmark{3},
S.~Weimann\altaffilmark{3},
S.~Garrappa\altaffilmark{3},
Y.~Shanni\altaffilmark{1},
P.~Chen\altaffilmark{1},
E.~Zimmerman\altaffilmark{1}
}

\altaffiltext{1}{Department of particle physics and astrophysics, Weizmann Institute of Science, 76100 Rehovot, Israel.}
\altaffiltext{2}{University of California, Berkeley, Department of Astronomy, Berkeley, CA 94720}
\altaffiltext{3}{Faculty of Physics and Astronomy, Astronomical Institute (AIRUB), Ruhr University Bochum, 44780 Bochum, Germany}

\begin{abstract}


The Large Array Survey Telescope (LAST) 
is a wide-field telescope designed to explore the variable and transient sky with a high cadence
and to be a test-bed for cost-effective telescope design.
A LAST node is composed of 48 (32 already deployed), 28-cm f/2.2 telescopes.
A single telescope has a 7.4\,deg$^{2}$ field of view and reaches a
$5\sigma$ limiting magnitude of 19.6 (21.0) in 20 ($20\times20$) seconds (filter-less),
while the entire system provides a 355\,deg$^{2}$ field of view.
The basic strategy of LAST is to 
obtain multiple 20-s consecutive exposures of each field (a visit).
Each telescope carries a 61\,Mpix camera,
and the system produces, on average, about 2.2\,Gbit\,s$^{-1}$.
This high data rate is analyzed in near real-time
at the observatory site, using limited computing resources (about 700 cores).
Given this high data rate, 
we have developed a new, efficient data reduction and analysis pipeline.
The LAST data pipeline includes two major parts:
(i) Processing and calibration of single images,
followed by a coaddition of the visit's exposures.
(ii) Building the reference images and performing image subtraction and transient detection.
Here we describe in detail the first part of the pipeline.
Among the products of this pipeline are
photometrically and astrometrically calibrated single 
and coadded images,
32-bit mask images marking a wide variety of problems and states of each pixel,
source catalogs built from individual and coadded images, 
Point Spread Function (PSF) photometry, merged source catalogs, 
proper motion and variability indicators,
minor planets detection, calibrated light curves,
and matching with external catalogs.
The entire pipeline code is made public.
Finally, we demonstrate the pipeline performance on real data taken by LAST.

\end{abstract}


\section{Introduction}
\label{sec:Introduction}

Sky surveys produce a large amount of data.
Therefore, efficient and successful use of these datasets
requires a diversity of high-quality data products that enable
achieving the science goals in a timely manner.

Here we describe the first-step pipeline (out of two)
developed for the Large Array Survey Telescope (LAST) and the {\it ULTRASAT} space mission (\citealt{Sagiv+2014_ULTRASAT, Shvartzvald+2023PASP_ULTRASAT_Overview}).
LAST is a new, cost-effective sky survey (\citealt{Ofek+BenAmi2020_Grasp_SkySurvrys_CostEffectivness}).
A LAST node contains 48, 28-cm f/2.2 telescopes.
Each telescope carries a 61\,Mpix
(about 6400 by 9600 pixels)
backside-illuminated CMOS detector with $3.76\,\mu$m pixels.
This yields a pixel scale of $1.25''$\,pix$^{-1}$ 
and a 7.4\,deg$^{2}$ field of view (FOV) per telescope.
The LAST system design and overview are discussed in \cite{Ofek+2023PASP_LAST_Overview},
while the LAST science goals are reviewed in \cite{BenAmi+2023PASP_LAST_Science}.
Among the LAST main science goals are the search for optical counterparts of gravitational-wave events
(e.g., \citealt{Abbott+2017_GW170817_LIGO}), exoplanets, supernovae, fast transients
(e.g., \citealt{Drout+2014_Rapidly_Evolving,Ho+2021_ZTF_RapidlyEvolvingTransients,Ofek+2021_AT2018lqh_RapidlyEvolving}), and the study Solar System objects.
Some LAST science results are presented in  Ofek et al. (in prep.),
and Ho et al. (submitted).
The LAST default observing strategy includes
observing each field for 5 to 20 successive 20\,s exposure images
(a visit).
This strategy has several unique advantages:
(i) It allows us to screen for satellite glints (\citealt{Corbett+2020_SatellitesGlints}, 
\citealt{Nir+2020_Satellites_Glints_FlaresLimit}, \citealt{Nir+2021_RNASS_GN-z11-Flash_SatelliteGlint})
which appear as point sources in a single image (due to their sub-second duration);
(ii) It allows the detection of main-belt asteroids due to their motion within a single visit (400\,s);
(iii) It enables us to explore stellar variability on time scales as short as 20~s, which is relevant for exoplanets transiting white dwarfs, and for flare stars (see \citealt{BenAmi+2023PASP_LAST_Science}).

With a nominal exposure time of 20\,s and dead time between telescope visits, 
LAST produces data at a rate of 2.2\,Gbit\,s$^{-1}$.
This data rate is about 70\% higher than
the expected LSST data rate (assuming a 30\,s exposures; \citealt{Ivezic+2019_LSST_Survey}).
Such a high data rate requires either considerable computational resources or a highly efficient 
data reduction pipeline.
The LAST data reduction pipeline consists of two parts.
The first part is responsible for processing raw images and producing calibrated images, including coadded images of successive
snapshots of a single visit.
The second part is responsible for performing image subtraction and transient detection.
In this paper, we describe the first part of the pipeline,
while the second part will be described in a following publication.

In \S\ref{sec:Overview} we provide an overview of the main pipeline steps.
The software package is discussed in \S\ref{sec:Software},
while in \S\ref{sec:DetailedDescription} we provide a detailed description of the steps
outlined in \S\ref{sec:Overview}.
In \S\ref{sec:DataProducts} we discuss the main data products,
and in \S\ref{sec:Performances} we describe the pipeline's on-sky performance.
In \S\ref{sec:Timing} we analyze the pipeline speed performance , 
and 
\S\ref{sec:Conclusion} summarizes our conclusions.

\section{Pipeline overview}
\label{sec:Overview}

Here we briefly describe the high-level data processing steps implemented in the LAST pipeline.
The following steps 
are carried out for each science image that is taken
during the night
and are fully addressed in \S\ref{sec:DetailedDescription}
Apart from the science image data, the processing requires bias and flat-field images
along with their corresponding bit mask images. The flat-field images are produced once per night.

\begin{enumerate}
    \item A 32-bit mask image is created for the science image and the saturated pixels are flagged therein. The mask image is further populated in the following steps (see Table~\ref{tab:PixLevelBitMask}).
    \item A master dark image is subtracted from the raw science image (including the darkened/overscan region). The bit mask of the dark image is propagated to the science image bitmask using the {\it or} operator.
    \item The residual dark value of the global overscan/darkened pixels is subtracted from the image.
    \item A non-linearity correction is applied.
    \item The science image is divided by the most recent flat field image, and the bit mask of the flat field image is propagated to the science image bit mask.
    \item The darkened (analogously to overscan) pixels are trimmed.
    \item The image is interpolated over saturated-pixels, NaN pixels, and negative value pixels. Each of the interpolated pixels is specifically flagged in the bit mask.
    \item The pixel values are multiplied by the gain. 
    \item The image is partitioned into 24 sub-images of $1726\times 1726$-pixel size,
    including a $\gtrsim$~64 pixels overlap between the sub-images (to avoid losing sources that fall near the edges).\\
    \\
    After this step, all processing steps are done on the sub-images.\\
    \item Background and variance images are estimated for each image. 
    \item Matched-filtering\footnote{Matched filtering in independent and identically distributed Gaussian noise is given by cross-correlation of the image with the filter (e.g., \citealt{Zackay+2017_CoadditionI}).} of the background-subtracted images with a template bank of Gaussian filters with width (sigma) of 0.1, 1.0, 1.5, 2.6, and 5.0 pixels is performed. 
    \item The match-filtered images are searched for sources, and for each filter and each source the $S/N$ ratio is calculated and a convolution-based Point-Spread Function (PSF) photometry flux is measured. Hence, a source catalog for each sub-image is generated.
    \item The first and second moments are calculated for each of the cataloged sources. In addition, aperture photometry is performed (in three apertures) with the background and standard deviation measured in an annuli around the sources.
    \item{The PSF in each sub-image is estimated, and PSF photometry is performed.}
    \item The sources are classified into delta function (e.g., cosmic rays), and stellar PSF sources. Cosmic rays are removed from the catalog, and their positions are marked in the bit mask image.
    \item The bit mask information is propagated to the source catalog. 
    \item The astrometric solution is found for each sub-image.
    \item The source catalog is updated with the astrometric information.
    \item The sources in the (5 to) 20 successive images of a visit are matched, and a matched source catalog containing all detected sources is produced (for each sub-image).
    \item Proper motion and variability of each of the snapshot-matched sources are measured.
    \item A search for asteroids in the 20 successive images of a visit is performed.
    \item The visit-matched sources are matched with external catalogs.
    \item The 20 successive sub-images (with overlap) of the same position in the detector are transformed into the same reference frame and coadded. 
    \item A source catalog for each coadded sub-image is produced (similar to the catalogs of the individual images).
    \item The astrometry of the coadded catalog is refined using the PSF-fitted source positions.
    \item All the data products are saved.
\end{enumerate}

The second part of the pipeline (Ofek et al., in prep)
is responsible for the reference image coaddition, image subtraction, 
and detection of transient sources.

\section{The software package}
\label{sec:Software}

The LAST pipeline is a part of the {\tt AstroPack}/{\tt MAATv2} package 
(\citealt{Ofek2014_MAAT}) that is being developed for the image processing 
system of the {\sl ULTRASAT} 
(\citealt{Sagiv+2014_ULTRASAT, Shvartzvald+2023PASP_ULTRASAT_Overview}) space mission.
{\tt AstroPack}/{\tt MAATv2} is mainly written in MATLAB, with some tools
in {\tt C}, {\tt C++} and {\tt Python}.
The code consists of over $10^{5}$ lines, and we put emphasis on documentation 
and on the built-in testing suite (unittest).
Containing a wide variety of tools, the code is designed to be efficient, modular, and
flexible, with several goals in mind, including fast adaption for a wide range of 
pipelines and complex analysis of telescope data in a few steps.

By design, data (e.g., images, catalogs) is stored in several classes, which contain the basic functionality,
while the high-level functionality is available in other specialized functions or classes.
The object data containers are arrays, and therefore, most of the operations can be 
performed simultaneously on multiple instances of the objects.

Code efficiency is a very important requirement, as the LAST data have to be processed
with limited computational resources. Therefore, employment of an off-the-shelf code
was not an option (see \S\ref{sec:Timing}).

{\tt AstroPack} will be described in detail in a future publication (see also 
\citealt{Ofek2014_MAAT}). However, a certain amount of documentation has already been 
included in a freely distributed version available at GitHub\footnote{\url{https://github.com/EranOfek/AstroPack}}.
In addition, an online wiki help pages are available\footnote{\url{https://github.com/EranOfek/AstroPack/wiki}}.

\section{Detailed description of the pipeline}
\label{sec:DetailedDescription}

\subsection{Dark and Flat calibration}
\label{sec:DarkFlat}

A sequence of 20 dark images, each of a 20\,s exposure, are
obtained with each camera, every few weeks.
The dark images are combined\footnote{Using {\tt AstroPack} {\tt imProc.dark.bias}.} 
using a sigma-clipped mean with a 5-$\sigma$ rejection.
In the same step, we also calculate the variance image of the dark measurement,
and create a bit mask for the dark image. 
The dark image bit mask includes the following bits,
described in Table~\ref{tab:PixLevelBitMask}:
{\tt LowRN}, {\tt HighRN}, {\tt DarkHighVal}, {\tt DarkLowVal}, {\tt BiasFlaring}.

The flat images are generated on a nightly basis.
Currently, the flats are generated using twilight images (see overview paper).
Each flat image is normalized by its own image median.
Next, the normalized flat images are combined\footnote{Using {\tt AstroPack} {\tt imProc.flat.flat}.} using a sigma clipped median with a 5-$\sigma$ rejection.
The combined flat image is divided by its own median.
We also generate a variance image for the flat and a flat bit mask image.
The bit mask image includes the bits
{\tt FlatHighStd}, {\tt FlatLowVal}, {\tt NaN} (see Table~\ref{tab:PixLevelBitMask}).

\subsection{Bit masks}
\label{sec:BitMask}

For each calibrated science image, a 32-bit mask image is generated. The bit mask is 
designed to indicate a wide range of states and conditions in the image, and it is also 
propagated to the source catalog and to the coadded images.
In Table~\ref{tab:PixLevelBitMask} we list the LAST bit dictionary, with the bit 
definitions.
Some of the bits are measured from the raw image itself (e.g., {\tt Saturated}), while 
some are propagated from other calibration images (e.g., {\tt LowRN}).
Population of some of the bits (i.e., {\tt Ghost}) is not yet implemented in the pipeline,
and some are required for other missions (e.g., {\sl ULTRASAT}; \citealt{Sagiv+2014_ULTRASAT}, Shvartzvald et al.,  in prep.).
\begin{deluxetable*}{llll}
\tablecolumns{4}
\tablewidth{0pt}
\tablecaption{Pixel-Level Bit Mask Dictionary}
\tablehead{
\colhead{Bit name}    &
\colhead{Index}   &
\colhead{Used}    &
\colhead{Description} \\
\colhead{}       &
\colhead{}       &
\colhead{}       &
\colhead{}  
}
\startdata
Saturated        & 0 & Y & Pixel is saturated (value$>$62000)\\
LowRN            & 1 & Y & Pixel noise is low (Var$<$median(Var)$\times0.05$) (probably dead) \\
HighRN           & 2 & Y & Pixel noise is high (Var$>10\times$median(Var)) (noisy pixel) \\
DarkHighVal      & 3 & Y & Pixel bias/dark value is high ($>2\times$Mean) \\
DarkLowVal       & 4 & Y & Pixel bias/dark value is low ($<0.3\times$Mean) \\
BiasFlaring      & 5 & Y & Pixel dark level is possibly flaring (one of the values $>$20-sigma) \\
NaN              & 6 & Y & Pixel is NaN due to illegal arithmetic \\
FlatHighStd      & 7 & Y & Flat StD/$\sqrt{N}$ is high (Std/$\sqrt{N}>$0.01) \\
FlatLowVal       & 8 & Y & Flat normalized value is low ($<$0.1) \\
LowQE            & 9 & N & Pixel with low QE ($<$0.5 from the nominal value) \\
Negative         & 10 & Y & Negative pixel after processing \\
Interpolated     & 11 & Y & Pixel with an interpolated value \\
Hole             & 12 & Y & Hole (anti-star) \\
Spike            & 13 & Y & Diffraction spike \\
CR\_DeltaHT      & 14 & Y & Cosmic Ray (CR) identified using hypothesis testing to delta function \\
CR\_Laplacian    & 15 & N & CR identified using Laplacian filter \\
CR\_Streak       & 16 & N & CR streak \\
Ghost            & 17 & N & Ghost \\
Persistent       & 18 & N & Persistent charge \\
Xtalk            & 19 & N & Cross talk \\
Streak           & 20 & N & Streak \\
ColumnLow        & 21 & N & Bad column low values \\
ColumnHigh       & 22 & N & Bad column high values \\
NearEdge         & 23 & Y & Near image edge \\
NonLinear        & 24 & Y & Non linear \\
Bleeding         & 25 & N & Possible bleeding/blooming \\
Overlap          & 26 & Y & In an overlap region \\
SrcNoiseDominated& 27 & Y & S/N is source-dominated (Val$-$Back)$>$Back \\
GainHigh         & 28 & N & In dual-gain detectors: the gain is high \\
CoaddLessImages  & 29 & Y & Number of images in coadd is less than 60\% of the requested \\
SrcDetected      & 30 & N & Source detected (all the pixels within the 1$\times$FWHM radius from a detected source) 
\enddata
\tablecomments{This bit-mask dictionary is used in all bit mask images and {\tt FLAGS} 
columns in source Tables. The Used column indicates if the bit is used in the current 
version of the pipeline. Some bits are not used because they are not relevant for LAST 
(e.g., {\tt GainHigh}), or we have not yet developed the required algorithm (e.g., {\tt 
Ghost}). The {\tt SrcNoiseDominated} bit is populated only in the Coadd image.
An order of magnitude estimate, based on several examples, of the fraction of pixels with
the {\tt LowRN}, {\tt HighRN}, {\tt DarkHighVal}, {\tt DarkLowVal},
{\tt BiasFlaring}, and {\tt CR\_DeltaHT} bits {\it on} are about:
$1.6\times10^{-5}$,
$0.0009$,
$0.0008$,
$0.0002$,
$0.0009$,
and
$2\times10^{-5}$, respectively.}
\label{tab:PixLevelBitMask}
\end{deluxetable*}

\subsection{Basic calibration}
\label{sec:BasicCalib}

The basic calibration step\footnote{Carried out using the {\tt CalibImages} class, {\tt processImages} method.}
includes populating the {\tt Saturated}, and
{\tt NonLinear}
bits,
subtracting a dark image, subtracting a global median of the darkened pixels, 
dividing by the flat image, cropping the darkened pixel region,
and multiplying the image by its gain, such that the effective gain is 1.
The original detector gain is stored in the {\tt ORIGGAIN} header keyword.
The non-linearity correction is applied
before the division by the flat image.
The non-linearity of each camera is measured in the lab,
and stored in a configuration file.
Measurements of the non-linearity are presented in \cite{Ofek+2023PASP_LAST_Overview}.

Next, the image is partitioned into 24 sub-images, sized $1726\times1726$ pixels.
The sub-images contain a $\gtorder64$-pixel overlap region.
The overlap region is marked in the bit mask image using the {\tt Overlap} bit.
In addition, pixels within 10\,pixels from the image edge are marked in the bit-mask using the {\tt NearEdge} bit.

There are several reasons behind image partitioning:
(i) Any image calibration process (e.g., astrometry and photometry), 
and steps related to image subtraction (e.g., registration and PSF estimation),
become more accurate when the field of view decreases;
(ii) Data transfer of small files is more practical for a wide range of applications;
(iii) The full image is too large to fit in the cache memory of most current computers.
Indeed, tests show that using small
sub-images improve the speed performance,
compared to full-size images;
(iv) With our astrometric solution approach (\S\ref{sec:Astrometry}), 
the astrometry speed performance is improved by about two orders of magnitudes.

\subsection{Background mean and variance estimation}
\label{sec:back}

Reliable background-mean and background-variance image estimations are crucial for source detection, 
photometry, and image subtraction. However, typically, it is not trivial to estimate 
them accurately and even 
their definition depends on their purpose.

Several methods of image background and variance estimation are implemented in the {\tt AstroPack} software.

The default strategy for LAST images is to estimate the background and variance in 
$128\times128$ pixel blocks\footnote{Implemented using {\tt imProc.background.background}.}.
In each block we estimate the mode and the variance.
The mode is estimated by calculating the histogram of all the pixels, 
with logarithmic bins that on average contain about 100 pixels,
and taking the highest value in the histogram.
The variance is estimated by calculating the width of the histogram that contains 
50\% of the data, and converting it to variance, assuming Gaussian statistics.
Next, the sparse (i.e., down-sampled) background and variance maps are interpolated and extrapolated
linearly to the full image.
We are currently working on improvements to this approach.

\subsection{Source detection and measurements}
\label{sec:SrcDet}

\begin{figure}
\includegraphics[width=8cm]{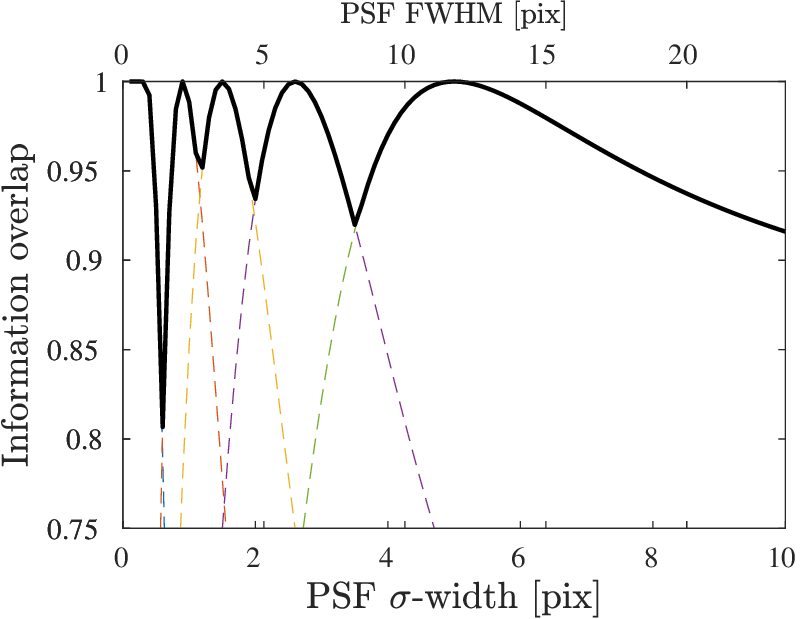}
\caption{The theoretical information-overlap of all the Gaussian filters we are using for source detection,
as a function of the source actual $\sigma$-width. The dashed lines are for individual filters (color-coded)
(i.e., 0.1, 1, 1.5, 2.6, and 5 pixels), while the solid line is for the maximum over the range.
The delta-function filter (of the $0.1$\,pix width) is shown by the peak on the left.
The information content using the formula in \cite{Zackay+2017_CoadditionI}.
The meaning of the information content loss is how well the filtering process can detect a source whose PSF is perturbed (i.e., different FWHM) compared with the filter PSF. The information loss is proportional to the variance and hence to the $(S/N)^{2}$.
\label{fig:PhotInformation}}
\end{figure}

The source detection algorithm is based on matched filtering with a template bank,
after subtracting the local background.
The default template bank includes five Gaussian filters with $\sigma$-width
of 0.1, 1, 1.5, 2.6, and 5 pixels.
The $0.1$\,pixel filter is supposed to mimic a delta function,
and it is used for cosmic ray detection via hypothesis testing (e.g., \citealt{Zackay+2016_ZOGY_ImageSubtraction}).
The $1.0$ and $1.5$ pixel filters are for (stellar) point-like sources, while the wider filters are appropriate to cover seeing variations, detection of extended sources, and optical ghosts.
Figure~\ref{fig:PhotInformation} presents the theoretical information overlap of all the filters we are using as a function of a source $\sigma$-width. The information loss is less than 10\% for sources with FWHM in the range of 1.7 to 16\,pixels ($2.1''$ to $20.6''$).

The matched-filter map is normalized by the square root of the filtered-image variance\footnote{In practice, in order to estimate the StD of the filtered image we measure the StD of the unfiltered images (which is not highly correlated) and multiply this StD by $\sqrt{\sum_{q}{P_{q}^{2}}}$, where $P_{q}$ is the value of the unit-normalized PSF at position $q$.}.
Next, the pixel-by-pixel maximum over all 
matched-filtered images is obtained,
and we search for local maxima in the combined matched-filter image\footnote{using the {\tt imUtil.sources.findLocalMax} function.}.
Any peak, in one of the five matched filter images, is declared as a source candidate,
if its highest value is above a threshold level of 5~$\sigma$.
We note that we normalize the matched filter image by the StD of the filtered
image\footnote{Unlike {\tt SExtractor} (\citealt{Bertin+1996_SExtractor}) 
that normalizes the matched-filter image by the StD of the unfiltered image. 
This causes the {\tt SExtractor} detection threshold not to be in units of sigma.}
such that the resulting image is in units of standard deviation
(e.g., \citealt{Zackay+2017_CoadditionI}).

Given a list of integer-pixel positions 
of possible sources and their $S/N$ using each 
of the filters, we calculate the Gaussian-weighted first and second moments for each source. 
Next, a convolution-based PSF photometry of the source candidates is performed. 
This is done by normalizing the filtered image to flux units
(see the formulae in \citealt{Zackay+2017_CoadditionI}), 
and reading the flux value at the whole pixel position of the source.
Aperture photometry is done in three apertures of 2, 4, and 6-pixel radius.
The aperture photometry is calculated by FFT-shifting
the sources to the pixel origin\footnote{The PSFs stamp size in each axis is uneven and the PSF center is in the origin of the central pixel.}, and summing all the pixels within some radius.
Although the apertures have a pixelated shape,
thanks to the FFT-shift it is consistent 
(i.e., has the same shape, relative to the star position) for all the sources.
In addition, for each sub-image, we also estimate the PSF and use it to perform
PSF-fit photometry (see \S\ref{sec:PSFphot}).
One disadvantage of this approach is that in crowded regions it overestimates the background and variance, and therefore misses real sources.
To correct this, in the future, we will replace this routine with an iterative PSF measurement and subtraction similar to the approach implemented in DAOphot (\citealt{Stetson1987_DAOPHOT}).

The first and second source moments are calculated using an iterative algorithm.
In the first iteration, we assume a flat weight function, and then we narrow down
the weight function (a Gaussian of some finite width) from iteration to iteration,
until the $\sigma$-width of the Gaussian weight function reaches 1.5-pixels at the final iteration.
Our experimentation suggests that this algorithm converges faster, 
compared to a non-variable weight function, and results in
more accurate estimations\footnote{Implemented in 
{\tt AstroSpec imUtil.image.moment2}.}.
A full list of the columns listed in
the images catalog is presented in 
Table~\ref{tab:IndivImagePhot}.
\begin{deluxetable*}{lll}
\tablecolumns{3}
\tablewidth{0pt}
\tablecaption{Individual (epoch) image source catalog columns}
\tablehead{
\colhead{Column Name}    &
\colhead{Units}    &
\colhead{Description} \\
\colhead{}       &
\colhead{}       &
\colhead{}  
}
\startdata
RA                   & deg & J2000.0 Right Ascension \\
Dec                  & deg & J2000.0 Declination \\
XPEAK                & pix & Matched filter peak whole pixel $X$ position \\
YPEAK                & pix & Matched filter peak whole pixel $Y$ position \\
X                    & pix & First moment pixel $X$ position \\
Y                    & pix & First moment pixel $Y$ position \\
X2                   & pix$^2$ & Second moment $X^2$ \\
Y2                   & pix$^2$ & Second moment $Y^2$ \\
XY                   & pix$^2$ & Second moment $XY$ \\
SN\_$i$               & & S/N measured for Gaussian matched filter with $\sigma$ of $0.1$, $1.0$, $1.5$, $2.6$, and $5.0$\,pix ($i\in 1..5$) \\

BACK\_IM             & e$^{-}$& The background level in the background image at the source position \\
VAR\_IM              & e$^{-}$& The variance level in the variance image at the source position\\
BACK\_ANNULUS        &e$^{-}$ & The background as estimated as the median in an 8 to 12 pix annulus around the source.\\
STD\_ANNULUS         & & The StD in an 8 to 12 pix annulus around the source.\\

FLUX\_APER\_$i$      &e$^{-}$ & Aperture photometry instrumental flux, for aperture radii of 2,4,6\,pixels ($i\in 1..3$)\\ 
FLUXERR\_APER\_$i$      &e$^{-}$ & Aperture photometry instrumental flux error, corresponding to FLUX\_APER\_$i$ \\ 

MAG\_APER\_$i$      &e$^{-}$ & Aperture photometry calibrated magnitude (Luptitude), for aperture radii of 2,4,6\,pixels ($i\in 1..3$)\\ 
MAGERR\_APER\_$i$      &e$^{-}$ & Aperture photometry magnitude error, corresponding to MAG\_APER\_$i$ \\ 

FLUX\_CONV\_$i$  &e$^{-}$ & Instrumental flux measured from PSF convolution (with the same $\sigma$ as in SN$_{i}$ ($i\in 1..5$) \\

MAG\_CONV\_$i$ & mag & Calibrated Mag (Luptitude) corresponding to FLUX\_PSFCONV\_$i$\\
MAGERR\_CONV\_$i$ & mag & Calibrated Mag error corresponding to MAG\_PSFCONV\_$i$ \\

MAG\_PSF & mag & Calibrated PSF-fit Mag (Luptitude)\\
MAGERR\_PSF & mag & Calibrated PSF-fit Mag error\\

FLAGS            & & Bit-mask flags propagated from the bit-mask image {\it or} over bits within a 7$\times$7 pixel box around the source\\
MergedCatMask           &     & Bit mask indicating association with external matched catalogs (Table~\ref{tab:ExternalCatalogs}) 
\enddata
\tablecomments{All magnitudes are calculated assuming fiducial color of $B_{\rm p}-R_{\rm p}=0$.
The MergedCatMask column is available only in the coadded image and in the merged visit catalog.
\label{tab:IndivImagePhot}}
\end{deluxetable*}

\subsection{PSF-fit photometry}
\label{sec:PSFphot}

For each sub-image we also estimate the PSF and fit it to all the sources.
The PSF is estimated by selecting isolated sources with $500>S/N>30$. 
We cut small stamps around each source, and 
all the source stamps are aligned using a sub-pixel fft shift\footnote{Using {\tt imUtil.trans.shift\_fft}.}, and coadded using a sigma-clip mean, with 5-$\sigma$ rejection\footnote{Using imProc.psf.constructPSF}.
The wings of the pixelated PSF are smoothed by forcing the PSF wings 
above a radius of 5 pixels to decay exponentially\footnote{It is critical that the PSF wings will be noiseless, otherwise in the image subtraction step a convolution of the PSF with bright sources  will introduce new fake sources.}.

Next, we perform a PSF-fit photometry\footnote{Using {\tt imProc.sources.psfFitPhot}.} using the estimated PSF.
This is done using the following algorithm:
The first-moment position is used as the initial guess for the PSF position.
In each iteration, we calculate the two-dimensional gradient and second derivative 
of the $\chi^{2}$ around the current position,
and use this gradient and second derivative to extrapolate the position
of the local $\chi^{2}$ minimum. 
However, to avoid large jumps induced by noise,
we limit the maximum step size to about 0.1\,pixels.
When we calculate the $\chi^{2}$, for each source position,
the PSF flux is fitted as an independent parameter.
In the future, we plan to replace this routine with a multi-iteration PSF-fitting function,
that has better performance in crowded regions.

\subsection{Astrometry}
\label{sec:Astrometry}

The astrometry code is based on some general ideas from \cite{Ofek2019_Astrometry_Code}.
However, the code was rewritten for modularity and speed considerations (see also \citealt{Corbett+2022SPIE_ArgusArray_DataArcitecture}).
Specifically, we required improvements in speed performances by over an order of magnitude, compared to
other tools.
This is achieved by using two kinds of astrometry routines:
one for solving the astrometry from scratch, including pattern
matching\footnote{Using {\tt imProc.astrometry.astrometryCore}.},
and the second for refining an existing astrometric solution\footnote{Using {\tt imProc.astrometry.astrometryRefine}.}
Since pattern matching is relatively expensive, 
we run it for one or two sub-images of the first epoch in the visit, 
while for all other sub-images and epochs in the visit we refine the astrometry 
based on the solution extrapolated from the nearby sub-images.

In each sub-image we first fit an affine transformation,
and the residuals are fitted with a 3rd-order polynomial in X and Y.
Although second-order polynomials can fit the smooth atmospheric refraction 
to better than about 1\,mas over our sub-image field 
for an altitude $>30$\,deg above the horizon, 
we use higher (3rd) order polynomials in order to capture some of the distortions 
induced by the atmospheric scintillation (\citealt{Lindegren1980_AtmosphericScintilation_GroundBasedAstrometry, Shao+Colavita1992_AtmosphericScintilation_GroundBasedAstrometry, Ofek2019_Astrometry_Code}).

An additional feature of our code is the ability to query for the reference catalog once, 
and use the same reference catalog for all other 19 sub-images of the same visit.
Currently, the astrometry is done against the GAIA-DR3 catalog 
(\citealt{GAIA+2016_GAIA_mission}, \citealt{GAIA+2021_GAIAEDR3_Summary_Content}, \citealt{Lindegren+2021_GAIAED3_AstrometricSolution}),
and includes the sources' proper motions, and rejects sources with large astrometric residuals.

\subsection{Photometric calibration}
\label{sec:PhotCalib}

Each sub-image is photometrically calibrated against
the GAIA-DR3 catalog (\citealt{GAIA+2021_GAIAEDR3_Summary_Content}).
Currently, the calibration procedure is as follows.
We match the sources in each sub-image to the GAIA-DR3 catalog. 
Sources with measured photometric error better than 0.01\,mag, and $S/N<1000$ are selected.
Next, we fit the following equation
\begin{eqnarray}
    m_{s} - B_{{\rm p}, s} & = & \alpha_{1} + \alpha_{2} [(G_{s} - R_{{\rm p}, s}) - \overline{C}],
    \label{eq:ZPphot}
\end{eqnarray}
where $m_{s}$ is the instrumental magnitude for source $s$,
$G_{s}$ 
$B_{{\rm p}, s}$ 
$R_{{\rm p}, s}$ 
are the GAIA $G$, $B_{\rm p}$, and $R_{\rm p}$ AB magnitudes\footnote{GAIA-DR3 $B_{\rm p}$ Vega$-$AB magnitude is $-0.0155$\,mag.}, respectively, for source $s$,
$\overline{C}$ is the median color for sources in the field,
and $\alpha_{1}$ is the zero point 
and $\alpha_{2}$ is the color coefficient.
The fitted value of these free parameters, as well as $\overline{C}$, are stored in the header (see Table~\ref{tab:HeaderKeys}).
The photometric calibration is calculated using the
{\tt MAG\_PSF} magnitude\footnote{If not available then using the {\tt MAG\_APER\_3} magnitude insted}, and then it is applied to all the magnitudes (all the apertures, convolutions, and PSF photometry). 
The magnitude type used in the calibration process is indicated in the header keyword {\tt PH\_MAGT} (see Table~\ref{tab:HeaderKeys}).
This means that, excluding the {\tt MAG\_PSF}, our photometry may suffer
from aperture correction errors.
Currently, this is not corrected and is the responsibility of the user.
In future versions, we may correct for this effect.

When reporting the calibrated photometry in the catalog we apply the image zero point ($\alpha_{1}$),
but we set $\alpha_{2}$ to zero.
The reason for this is that the color is not necessarily known for all sources.
Therefore, the reported magnitudes are in the LAST {\it native AB system}.

Based on the photometric calibration,
we estimate the image limiting magnitude for source detection.
This is done by fitting the $\log_{10}(S/N)$ vs.~the star magnitude, 
and color, and reading the best-fit magnitude value 
at $S/N=5$ and $B_{\rm p}-R_{\rm p}=1$\,mag.
The limiting magnitude, as well as other parameters,
are added to the image headers (see Table~\ref{tab:HeaderKeys}).
We note that the fitted color terms are correlated with the radial distance from the field center.
 
A new method to perform the photometric calibration is being developed 
and will be included in future versions of the code.
 
\subsection{Interpolation over masked pixels}
\label{sec:interp}

In order to make sure that the coadd images are clean and to make the photometry more reliable, 
we interpolate over some pixels which have problematic bitmask values.
All the interpolated pixels are marked as such in the bit mask (using the {\tt Interpolated} bit).
The main rationale behind this step is that, first, any convolution operation will be less affected 
by sharp features (e.g., cosmic rays) that can cause long-range ringing artifacts, and,
second, photometric measurements will become more reliable.

The interpolation is done over pixels with one of the following bits {\it on}:
{\tt Saturated},
{\tt DarkHighVal,}
{\tt NaN},
and {\tt Negative}.
In the caption of Table~\ref{tab:PixLevelBitMask} we quote averages fractions of 
pixels with one of the bad pixel indicators.
The interpolation algorithm is described in \S\ref{sec:InterpAlgo}.

\begin{deluxetable*}{lll}
\tablecolumns{3}
\tablewidth{0pt}
\tablecaption{Special header keywords}
\tablehead{
\colhead{Keyword Name}    &
\colhead{Units}    &
\colhead{Description} \\
\colhead{}       & 
\colhead{}       &
\colhead{}  
}
\startdata
JD          & day   & UTC Julian day corresponding to the mid exposure time\\
\hline
PH\_ZP      & mag   & Photometric zero point ($\alpha_{1}$) \\
PH\_COL1    & mag   & First photometric color term ($\alpha_{2}$) \\
PH\_MEDC    & mag   & Median color ($\overline{C}$)\\
PH\_RMS     & mag   & Photometric fit rms (Equation~\ref{eq:ZPphot}) \\
PH\_NSRC    &       & Number of (bright) sources used in the photometric fit \\
PH\_MAGSY   &       & Magnitude system for magnitude quantities (default is native AB) \\
LIMMAG      & mag   & 5~$\sigma$ limiting magnitude for source detection in the calibrated (color and width corrected) magnitudes \\
BACKMAG     & mag\,arcsec$^{-2}$   & Background magnitude \\
PH\_MAGT    &  & Magnitude type (e.g., aperture/PSF) used for calibration\\
PH\_MAGTE   &   & Magnitude error type used for calibration\\
\hline
MEANBCK    & phot  & The mean value of the background image \\
MEDBCK     & phot  & The median value of the background image \\
STDBCK     & phot  & The StD value of the background image \\
MEANVAR    & phot  & The mean value of the variance image \\
MEDVAR     & phot  & The median value of the variance image \\
\hline
AST\_NSRC   &       & Number of sources used in the final iteration of the astrometric solution \\
AST\_ARMS   & arcsec& The astrometric asymptotic rms \\
AST\_ERRM   & arcsec& Theoretically estimated error of the reference system position \\
\hline
FWHM        & arcsec& FWHM as measured by the 50\% of the radial cumulative sum \\
MED\_A      & pix   & Median A (semi-major axis) of sigma of PSF \\
MED\_B      & pix   & Median B (semi-minor axis) of sigma of PSF \\
MED\_TH     & pix   & Median Theta (position angle) of the major axis of PSF \\
GM\_RATEX   & pix/s & Rate in global motion of 20 images in X direction \\
GM\_STDX    & pix   & StD of residuals of the linear fit of global motion in X direction \\
GM\_RATEY   & pix/s & Rate in global motion of 20 images in Y direction \\
GM\_STDY    & pix   & StD of residuals of the linear fit of global motion in Y direction 
\enddata
\tablecomments{Special image header keywords added during the image processing. 
The four blocks of the table refer to photometric, image background, astrometry, and image quality keywords (top to bottom).
{\tt AST\_ERRM} is calculated from the asymptotic rms divided by the square root of the number of sources ({\tt AST\_NSRC}). 
It is not the positional error, but rather the uncertainty in the knowledge of the coordinate system reference point compared to the GAIA/ICRS coordinate system.
The Global motion ({\tt GM\_}) keywords are added only for the coadd images.
FWHM, MED\_A, MED\_B, MED\_TH are populated only in the coadd images.
}
\label{tab:HeaderKeys}
\end{deluxetable*}

\subsection{Merged catalogs: 20 images light and position curves}
\label{sec:MergedCat}

The sources in each group of 20 catalogs of each sub-image in a single visit are matched using the following scheme:
The sources in the $i$ $(\in 2..20)$-image are matched against the first image sources with a 3-arcsec search radius.
Sources in the $i$-th image that have no counterparts are added to
the first image source list.
The result is a master list of sources that appear in at least one image.

Next, each source in this list is matched against all the sources in all the epochs of the visit, again with a 3-arcsec search radius.
For each property of interest (e.g., Right Ascension, Magnitude; see Table~\ref{tab:MergedCatMat}), we populate a matrix of all the measurements in all the epochs and for all the sources. Properties that are not available for some source/epoch are set to NaN.
These matrices are described in 
Table~\ref{tab:MergedCatMat}
and are saved in HDF5 files.

In addition, a matched sources catalog
is saved (in a FITS binary table)
with a row per matched source.
Also calculated and added to this catalog are proper motion fits (\S\ref{sec:Asteroids} and \S\ref{sec:PM}),
and a variability estimator for all the sources (see \S\ref{sec:var}).
The matched sources table columns are described in Table~\ref{tab:MergedCat}.

\begin{deluxetable*}{lll}
\tablecolumns{3}
\tablewidth{0pt}
\tablecaption{Merged catalog columns}
\tablehead{
\colhead{Column Name}    &
\colhead{Units}    &
\colhead{Description} \\
\colhead{}       &
\colhead{}       &
\colhead{}  
}
\startdata
RA                   & deg & J2000.0 Right Ascension \\
Dec                  & deg & J2000.0 Declination \\
Nobs                 &     & Number of observations used in proper motion fit\\
Noutlier             &     & Number of outliers (above $1.5''$) in proper motion fit\\
StdRA                & deg & R.A. StD of residuals from proper motion fit \\
StdDec               & deg & Dec. StD of residuals from proper motion fit \\
PM\_RA                & deg/day & R.A. proper motion \\
PM\_Dec               & deg/day & Dec. proper motion \\
PM\_TdistProb         & & Proper motion student-t distribution probability to reject the null hypothesis \\
FLAGS                & & {\it or} over all bit-mask of sources in epochs\\
Mean\_RA              & deg & Mean over all R.A. measurements \\
Std\_RA               & deg & StD over all R.A. measurements \\
Mean\_Dec             & deg & Mean over all Dec. measurements \\
Std\_Dec              & deg & StD over all Dec. measurements \\
Mean\_X               & deg & Mean over all X measurements \\
Std\_X                & deg & StD over all X measurements \\
Mean\_Y               & deg & Mean over all Y measurements \\
Std\_Y                & deg & StD over all Y measurements \\
Mean\_MAG\_CONV\_$i$    & mag & Mean over all MAG\_CONV\_$i$ measurements ($i\in{2..3}$)\\
Median\_MAG\_CONV\_$i$  & mag & Median over all MAG\_CONV\_$i$ measurements ($i\in{2..3}$)\\
Std\_MAG\_CONV\_$i$     & mag & StD over all MAG\_CONV\_$i$ measurements ($i\in{2..3}$)\\
RStd\_MAG\_CONV\_$i$    & mag & Robust StD over all MAG\_CONV\_$i$ measurements ($i\in{2..3}$)\\
Range\_MAG\_CONV\_$i$   & mag & Range of all MAG\_CONV\_$i$ measurements ($i\in{2..3}$)\\
Min\_MAG\_CONV\_$i$     & mag & Min of all MAG\_CONV\_$i$ measurements ($i\in{2..3}$)\\
Max\_MAG\_CONV\_$i$     & mag & Max of all MAG\_CONV\_$i$ measurements ($i\in{2..3}$)\\
Nobs\_MAG\_CONV\_$i$    &     & Number of MAG\_CONV\_$i$ measurements ($i\in{2..3}$)\\
Mean\_SN\_$i$           &     & Mean over all S/N of Gaussian PSFs ($i\in{1..4}$) \\
Std\_SN\_$i$            &     & StD over all S/N of Gaussian PSFs ($i\in{1..4}$) \\
Mean\_BACK\_IM          & e$^{-}$ & Mean over all BACK\_IM measurements \\
Std\_BACK\_IM           & e$^{-}$ & StD over all BACK\_IM measurements \\
Mean\_VAR\_IM           & e$^{-}$ & Mean over all VAR\_IM measurements \\
Std\_VAR\_IM            & e$^{-}$ & StD over all VAR\_IM measurements \\
Mean\_BACK\_ANNULUS     & e$^{-}$ & Mean over all BACK\_ANNULUS measurements \\
Std\_BACK\_ANNULUS      & e$^{-}$ & StD over all BACK\_ANNULUS measurements \\
Mean\_STD\_ANNULUS      & e$^{-}$ & Mean over all STD\_ANNULUS measurements \\
Std\_STD\_ANNULUS       & e$^{-}$ & StD over all STD\_ANNULUS measurements \\
Epoch$j$\_MAG\_CONV\_$i$ & mag & MAG\_CONV\_$i$ from $j$-th epoch ($i\in{2..3}$) \\ 
StdPoly                 & mag & StD of MAG\_CONV\_2 light curve after removing 5-th order polynomial \\
PolyDeltaChi2           &     & $\Delta\chi^{2}$ between 0 and 5-th order polynomial hypotheses \\
MergedCatMask           &     & Bit mask indicating association with external matched catalogs (Table~\ref{tab:ExternalCatalogs}) \\
LinkedAsteroid          &     & Asteroid index
\enddata
\tablecomments{There is one merged catalog table per sub-image per visit. 
This table contains a row per source with information gathered from all the epochs in a visit.
In column names starting with Epoch, the epoch index $j$ is saved with 3 digits (zero padded). 
Asteroid index - is a number indicating if a source was identified as an asteroid using the proper motion fit. 
NaN - indicates, not an asteroid. A positive index indicates an asteroid that was linked to another source in the merged catalog and all the linked sources have the same index. Negative numbers indicate a possible asteroid that is not linked (i.e., appears as a single proper motion fit).
The magnitudes in the merged catalog are the calibrated magnitude (i.e., no relative photometry is applied).
\label{tab:MergedCat}}
\end{deluxetable*}

\begin{deluxetable*}{lll}
\tablecolumns{3}
\tablewidth{0pt}
\tablecaption{Merged catalog matrices}
\tablehead{
\colhead{Dataset Name}    &
\colhead{Units}    &
\colhead{Description} \\
\colhead{}       &
\colhead{}       &
\colhead{}  
}
\startdata
RA                   & deg & J2000.0 Right Ascension \\
Dec                  & deg & J2000.0 Declination \\
X1                   & pix & First moment source X position \\
Y1                   & pix & First moment source Y position \\
SN\_$i$              &     & S/N of Gaussian PSF ($i\in{1..4}$) \\
MAG\_PSF             & mag & PSF magnitude measurements \\
PSF\_CHI2DOF         &     & $\chi^{2}$ per degree of freedom of PSF fit\\
MAG\_CONV$_i$        & mag & MAG\_CONV$_i$ measurements ($i\in{2..3}$)\\
MAGERR\_CONV$_i$     & mag & MAGERR\_CONV$_i$ measurements ($i\in{2..3}$)\\
MAG\_APER$_i$        & mag & MAG\_APER$_i$ measurements ($i\in{2..3}$)\\
MAGERR\_APER$_i$     & mag & MAGERR\_APER$_i$ measurements ($i\in{2..3}$)\\
FLUX\_APER$_i$       & mag & FLUX\_APER$_i$ measurements ($i\in{3}$)\\
FLAGS                & & {\it or} over all bit-mask of sources in epochs\\
BACK\_IM             & $e^{-}$ & Background at source position from background image\\
VAR\_IM              & $e^{-}$ & Variance at source position from background image\\
BACK\_ANNULUS        & $e^{-}$ & Background from annulus around source\\
STD\_ANNULUS         & $e^{-}$ & StD of background from annulus around source
\enddata
\tablecomments{List of datasets in the HDF5 files containing the merged catalog matrices.
Each dataset contains one merged matrix
of epoch vs.~sources,
for one measured property
(e.g., Right Ascension, Magnitude).
There is one merged matrices HDF5 file per sub-image per visit. 
Unlike in all the other data products, all the magnitudes
in this data products are relative photometry calibrated (\S\ref{sec:visitLC}),
after the photometric zero point calibration (see \S\ref{sec:visitLC}).
The relative photometry is done using the {\tt MAG\_APER\_3} magnitudes,
and the relative zero points are applied to all other magnitudes (ignoring aperture corrections).
\label{tab:MergedCatMat}}
\end{deluxetable*}

\subsection{Asteroids search}
\label{sec:Asteroids}

Before we describe specific channels for detecting asteroids, we briefly review our overall strategy.
We search for asteroids in the LAST images using several methods.
each method is designed for asteroids with different angular speeds:

(i) Asteroids moving faster than about $1''$\,s$^{-1}$, will be streaked and their $S/N$ will be diluted by a factor of $\gtorder5$.
These asteroids are best identified using streak detection methods.
Even with the fast and optimal algorithm presented in \cite{Nir+2018_StreakDetection},
streak detection is resource-expensive,
and we are currently developing a more efficient approach.

(ii) Asteroids having angular speeds below $1''$\,s$^{-1}$ and above $\approx0.06''$\,s$^{-1}$  
will be identified as matched sources (\S\ref{sec:MergedCat}) with less than four appearances.
Such asteroids will lack proper motion measurements
in the matched source tables (Table~\ref{tab:MergedCat}).
We detect these asteroids by fitting a linear motion to objects in the matched sources catalog
that have no proper motion fits.
Such objects are named orphans
(i.e., detected and matched in 1 to 3 exposures, within $3''$).
The search algorithm fits 2-D linear motion between all possible pairs of orphans.
To do this efficiently, we populate a k-d tree (e.g., \citealt{Press+2002_Book_NumericalRecepies})
with the four parameters of the fitted linear motion.
Finally, we search for orphan sources that have common linear motion parameters.
This algorithm\footnote{Implemented in {\tt AstroPack} {\tt imUtil.asteroids.pairsMotionMatchKDTree}.} 
(similar to \citealt{Denneau+2013PASP_PS1_MOPS_AsteroidsProcessing})
is efficient and its run time scales like $N^{2}$, where $N$ is the number of orphans.
On a single processor, it takes about 1\,s to match about 1000 orphan sources.
This functionality is being tested and will be incorporated into the pipeline
in the future.

(iii) With our typical astrometric precision (see \S\ref{sec:Perf_astrometry}), 
asteroids with angular speed between about $0.06''\,{\rm s}^{-1}$ and $3\times10^{-4}\,''\,{\rm s}^{-1}$, 
will be detected by the proper motion fit to the visit exposures.
This method covers a large fraction of the main-belt asteroids and is described in \S\ref{sec:PM}.

(iv) Asteroids with angular speeds slower than about $0.005''$/s will be a point source in the coadd images. 
A search for asteroids in the coaddition of 20 images can detect asteroids fainter by about 1.4\,mag, 
compared to asteroids detected in individual images.
These asteroids are searched for after image subtraction,
using the same approach as method (ii).

\begin{figure}
\includegraphics[width=8cm]{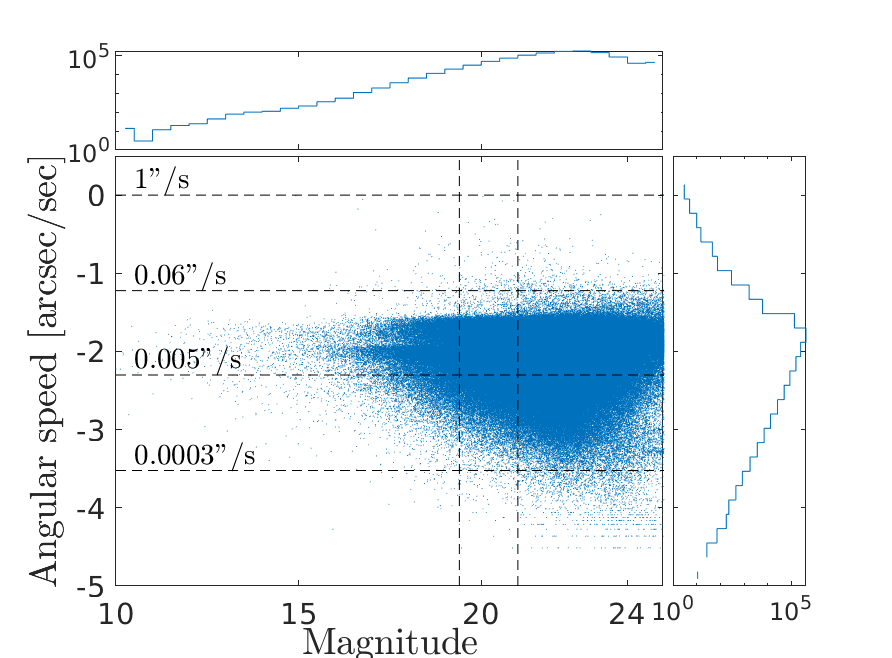}
\caption{The apparent magnitude vs. angular speed of all known numbered and unnumbered minor planets in some arbitrary epoch.
Also shown in horizontal lines are the angular speed regions discussed in \S\ref{sec:Asteroids}. The two vertical lines show the single exposure (19.4), and visit (21.0) limiting magnitude of LAST.
\label{fig:Asteroids_SpeedVsMag}}
\end{figure}

Figure~\ref{fig:Asteroids_SpeedVsMag} shows the apparent magnitude vs.~angular speed 
of all the currently known numbered and unnumbered minor planets\footnote{\url{http://ssd.jpl.nasa.gov/?sb_elem}}.
Also shown, in horizontal lines, are the angular speed regions discussed previously. 
The two vertical lines show the single exposure and visit limiting magnitude of LAST.

\subsection{Moving sources search using proper motion in the visit exposures}
\label{sec:PM}

The positions of each source that appears more than three times 
are fitted with two models: 
a stationary position hypothesis ($\mathcal{H}_{0, pm}$), 
and a linear proper motion hypothesis ($\mathcal{H}_{1, pm}$).
The positional errors are normalized (by a multiplication factor)
such that the $\chi^{2}$ per degree of freedom of $\mathcal{H}_{0, pm}$ is one.
Next, the Student's-t statistic
(assuming unknown StD) is calculated by
the proper motion value multiplied by the standard deviation of the number of epochs
and divided by the standard deviation of the residuals of the proper motion fit.
The Student's-t statistic is converted to the probability to reject the null hypothesis
(no motion) using the Student's-t distribution function.
Although, the probability to reject the null hypothesis should not depend on 
the number of observations, so we use
an adaptive threshold for finding asteroids.
The reason for this is that the number of trials
(look elsewhere effect) decreases when the number of epochs increases. 
In addition, an adaptive threshold is important
in order to deal with outliers. 
Specifically,
if the number of observations is above three, we require a false alarm probability of $10^{-4}$, 
while if the number of observations is larger than five, we require a false alarm probability
of $0.005$.

Other criteria include selecting sources, where the standard deviation of all the $S/N$ measurements is larger than $0.4$. 
The reason for this is that the expectation value of the StD of the $S/N$ is one,
and sources with abnormally low StD of $S/N$ usually indicate the presence
of bad pixels.
We also remove sources having one of the following bitmask values:
{\tt Saturated},
{\tt Spike},
{\tt CR\_DeltaHT},
{\tt CR\_Laplacian},
{\tt CR\_Streak},
{\tt Streak},
{\tt Ghost},
{\tt Persistent},
{\tt NearEdge},
and for sources with $S/N<8$, we also require that
the
{\tt DarkHighVal},
{\tt BiasFlaring},
{\tt FlatHighStd},
{\tt HighRN}
bits are off.

Next, moving sources are linked by matching their origin position
(i.e., with the fitted position at the visit's central epoch) to be consistent within 3\,arcsec.

\subsection{Relative photometry visit light curves}
\label{sec:visitLC}

An important feature of LAST is its
visit-observing strategy, where 
for each field $20\times20$\,s exposures are obtained, within a single 400\,s time window.
This, for example, allows for searching for exoplanets around about $10^{5}$ white dwarfs and for flare star studies (\citealt{BenAmi+2023PASP_LAST_Science}).
To achieve these goals, the 20 epochs light-curves are searched for variability.

We apply a relative photometry zero point.
In contrast to the analysis of the light curves generated over a larger number of epochs that
requires more sophisticated algorithms (e.g., \citealt{Ofek+2011_VLA5GHz_transient_search}),
here we use a simple and fast algorithm.
Specifically, we select all the sources that appear in all the epochs 
and have median photometric errors smaller than $0.02$\,mag.
We calculate the magnitude difference between the relative zero points of successive images.
We calculate the median of magnitude differences for all stars with mean magnitude error
smaller than $0.02$\,mag.
The differences between successive median-magnitude differences
are added to the images (skipping the 1st image).
The result is a magnitude system that is anchored to the photometric calibration
of the first image.
Next, we apply the zero points to all the calibrated magnitudes, 
to get the relative photometry magnitudes, for all the sources.
Note that the relative photometry zero points are calculated using only the 6-pixel radius aperture,
and are applied to all the photometry.
Therefore, aperture correction biases may exist, and the relative photometry may be further improved.
These relative photometry light curves are reported in the merged-catalogs matrices (Table~\ref{tab:MergedCatMat}),
while the photometry listed in the merged-catalogs table (Table~\ref{tab:MergedCat}) 
is before the relative photometry step (i.e., only photometric calibration is applied).

\subsection{Variability search}
\label{sec:var}

Based on the relative photometry visit light curves, 
we search for variable sources using several variability indicators:

(i) For each source we calculate the photometry $\chi^{2}$, relative to a constant flux model, and normalize the errors such that
the $\chi^{2}$ per degree of freedom is unity. Next, we fit a 5-th degree polynomial
and calculate the $\chi^{2}$ for this fit. We report in the merged catalog (Table~\ref{tab:MergedCat})
the $\Delta\chi^{2}$ between the 0-order polynomial fit $\chi^{2}$ and the 5-th order polynomial fit $\chi^{2}$
(i.e., {\tt PolyDeltaChi2} column).
(ii) For each source we calculate the StD and a robust StD.
In the future, we may consider adding a matched-filtering search based on
a predefined template bank.

\subsection{Cross matching with external catalogs}
\label{sec:ExternalCat}

The matched catalogs and coadd image catalogs are cross-matched with external catalogs.
To that end, we have generated a
{\tt catsHTM} catalog (\citealt{Soumagnac+Ofek2018_catsHTM}) of the catalogs listed in Table~\ref{tab:ExternalCatalogs}.
Also listed in this table, are the matched radius distance used for each catalog,
and a column containing a bit mask integer
indicating to which external catalog each line belongs.
For each source in the merged catalog (Table~\ref{tab:MergedCat}),
and in the coadd catalog, we provide a bit mask which is the {\it or} operation
between all the bit mask integers of all the matched external sources
that were matched to the source (i.e., {\tt MergedCatMask} column in Table~\ref{tab:MergedCat}).
\begin{deluxetable*}{llll}
\tablecolumns{4}
\tablewidth{0pt}
\tablecaption{External catalogs}
\tablehead{
\colhead{Catalog Name}    &
\colhead{Bit Mask}    &
\colhead{Search Radius} &
\colhead{Reference} \\
\colhead{}       &
\colhead{}       &
\colhead{[$''$]} &
\colhead{}
}
\startdata
GAIA-EDR3   & $2^0$    & 2  & \cite{GAIA+2016_GAIA_mission}; \cite{GAIA+2022yCat_GAIA_DR3_MainSourcesCatalog} \\
unWISE      & $2^1$    & 3  & \cite{2019ApJS_unWISE_catalog} \\
2MASS       & $2^2$    & 3  & \cite{Skrutskie+2006_2MASS} \\
GLADE$+$    & $2^3$    & 10 & \cite{Dalya+2018MNRAS_GLADE_GalaxyCatalog}\\
PGC         & $2^4$   & 10 & \cite{Paturel+1989_HyperLeda_PGC_Galaxy_Catalog} \\
SDSS-DR10   & $2^5$   & 2  & \cite{York+2000_SDSS_Summary}\\
PS1-DR1     & $2^6$   & 2  & \cite{Chambers+2016_PS1_Surveys} \\
DECaLS-DR4  & $2^7$  & 2  & \cite{Dey+2019AJ_DECaLS_Overview}\\
GALEX       & $2^8$  & 4  & \cite{Martin+2005ApJ_GALEX_Overview}\\
FIRST        & $2^9$ & 5  & \cite{Becker+1995_FIRST} \\
NVSS         & $2^{10}$ & 15 & \cite{Condon+1998_NVSS} \\
VLASS-epoch1 & $2^{11}$ & 5  & \cite{Lacy+2020PASP_VLASS_Science_SurveyDesign} \\
LAMOST-DR4   & $2^{12}$ & 2  & \cite{Luo+2018yCat_LAMOST_DR4_Online}\\
NED          & $2^{13}$ & 10 & \cite{Helou+1991_NED} \\
SDSS-spec-DR14 & $2^{14}$ & 10 & \cite{York+2000_SDSS_Summary}\\
ROSAT faint source  & $2^{15}$  & 30 & \cite{Zimmermann1994IAUC_ROSAT_SourceCatalog}\\
XMM           & $2^{16}$ & 10  & \cite{Webb+2022yCat_XMM_SerendipitousSourceCatalogue_4XMM_DR11} \\
ztfDR1var     & $2^{17}$  & 2 & \cite{Ofek+2020_ZTF_DR1_10milion_Variables} \\
WD            & $2^{18}$  & 2 & \cite{GentileFusillo+2021MNRAS_WD_Catalog_GAIAEDR3} \\
QSO           & $2^{19}$ & 2  &  \cite{Flesch2015_TheMilionQSO_Catalog} v7.2 \\
GAIA-DR3 extragalactic & $2^{20}$ & 10  & \cite{GAIA+2022yCat_GAIA_DR3_ExtragalacticCatalog} 
\enddata
\tablecomments{List of external catalogs matched to sources in the merged catalogs and coadd image catalogs. 
For each source, a 32 bit integer mask indicates (one bit per external catalog)
in which external catalog there is a possible counterpart of the source.
For each external catalog, a different search radius is adopted.
The GLADE$+$ version we use was last updated on 2022-Jul-10.
The NED redshift catalog was last updated on 2018-May-02.}
\label{tab:ExternalCatalogs}
\end{deluxetable*}

\subsection{20 images coadd and analysis}
\label{sec:coadd}

The successive exposures in a visit are registered using the images' World Coordinates System (WCS) and coadded.
The coaddition is done using a 5-sigma clipping,
with no weights and no background subtraction.
The rationale here is that, in the majority of cases, these 20 images
are affected by similar sky variance, and have similar PSFs. 
Therefore, the information loss is negligible (see \citealt{Zackay+2017_CoadditionI}, \citealt{Zackay+2017_CoadditionII}).
Next, the coadd image background and variance are estimated, sources are identified, and the astrometry is refined.
All these steps are similar to those described in \S\ref{sec:back}--\S\ref{sec:interp},
and \S\ref{sec:ExternalCat}.

\section{Additional details}
\label{sec:Details}
 
In this section, we several topics that, as far as we know,
are not standard procedures, and thus may be of special interest to the community.

\subsection{The flux-annulus-background bias}

\begin{figure}
\includegraphics[width=8cm]{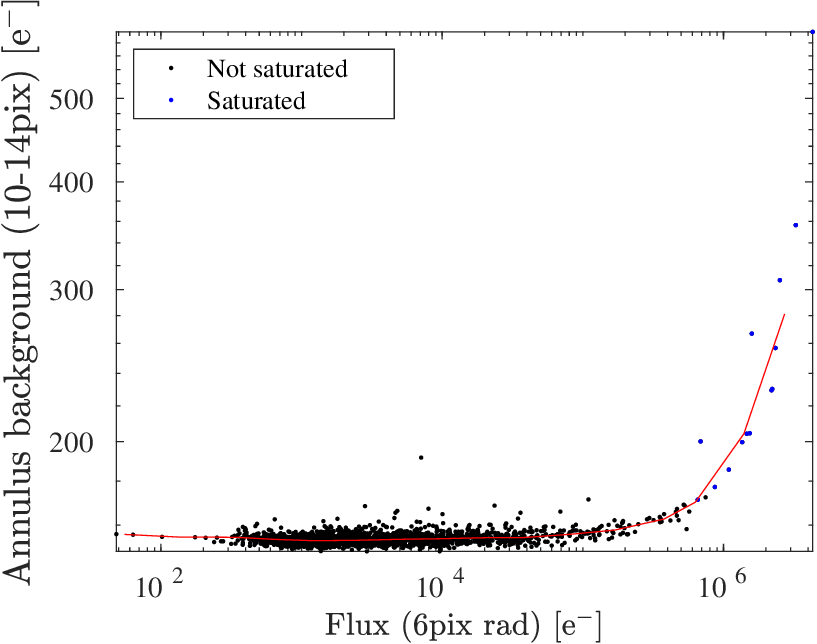}
\caption{The measured annulus background vs.~aperture photometry (in a 6\,pix radius) in a random LAST image. 
Saturated sources are marked as blue points.
\label{fig:Flux_AnnulusBack_bias}}
\end{figure}

\begin{figure}
\includegraphics[width=8cm]{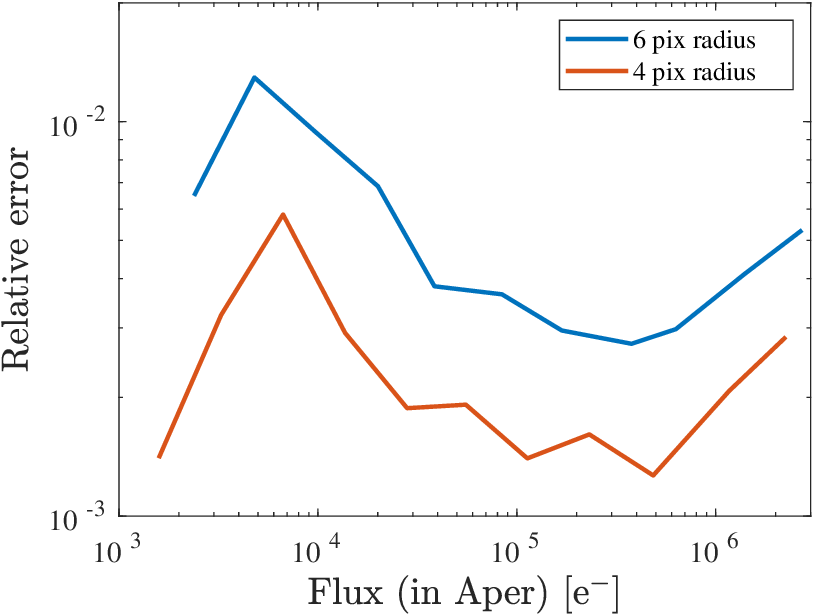}
\caption{The estimated relative error in aperture photometry due to the flux-annulus-background bias. 
Different line colors show different radii of aperture photometry.
\label{fig:Flux_AnnulusBack_biasRelErr}}
\end{figure}

The common procedure for estimating the local background and variance 
around sources is to measure these properties in an annulus of 
a constant radius, around each source.
In this case, the amount of light that leaks from a source into 
the annulus depends on the source flux.
In Figure~\ref{fig:Flux_AnnulusBack_bias} we present 
the measured annulus background vs.~aperture photometry (in a 6\,pix radius) 
in a random LAST image.
A clear bias in the background level is detected. 
This bias will tend to overestimate the background level around bright sources 
and therefore decrease their measured (background subtracted) flux.
The observed amplitude of this effect is larger than expected from a pure Gaussian PSF and may depend on
the quality of the optical system (aberrations and scattered light).
Figure~\ref{fig:Flux_AnnulusBack_biasRelErr} shows the estimated relative bias in photometry, as a function of flux,
for two apertures.

Currently, the photometry is not corrected for this effect, 
but offline tools are available in order to estimate the correction
as a function of flux and to apply it to data.
The estimator is generated by binning the flux vs.~annulus background distribution, in 0.3\,dex bins, and interpolating.

\subsection{Interpolation Algorithm}
\label{sec:InterpAlgo}

As described in \S\ref{sec:interp}, some pixels are interpolated over.
The interpolation algorithm is efficient and uses the following scheme.
We select a stamp around the interpolated pixel.
We prepare a Gaussian kernel, in which the central pixel is set to 0.
The modified Gaussian kernel is normalized to unity,
multiplied by the image stamp, and integrated over the stamp (i.e., local convolution).
The result of the local convolution replaces the value of the interpolated pixel\footnote{Implemented in {\tt imUtil.interp.interpImageConvPix}}.

\section{Data Products}
\label{sec:DataProducts}

In this section, the primary data products produced by the pipeline are described. 
Additional data products that are produced by the second (image subtraction) pipeline, 
and by additional post-processing (e.g., light curves) are not discussed here.
In \S\ref{sec:naming} we describe the naming conventions we use for all the data products.
Next, we describe the data products according to the following categories:
single exposures (\S\ref{sec:DP_IndivImages});
matched catalogs
(\S\ref{sec:DP_MatchedCat});
asteroids
(\S\ref{sec:DP_Asteroids});
and coadded images
(\S\ref{sec:DP_Coadd}).
The pipeline~I data products are summarized in Table~\ref{tab:DataProducts}.
\begin{deluxetable*}{llllll}
\tablecolumns{6}
\tablewidth{0pt}
\tablecaption{List of data products}
\tablehead{
\colhead{Level}    &
\colhead{Product}    &
\colhead{\#/vist/tel} &
\colhead{Availability} &
\colhead{File Type} &
\colhead{Comments} \\
\colhead{}       &
\colhead{}       &
\colhead{} &
\colhead{} &
\colhead{} &
\colhead{}
}
\startdata
Raw   & Image & 20  & $\sim60$\,days & FITS & Raw images\\
Proc  & Image & 480 & $\sim2$\,days & FITS & Processed epoch images \\
Proc  & Mask & 480 & $\sim2$\,days & FITS & Processed epoch mask images \\
Proc  & Catalog & 480 & $\infty$ & FITS & Catalogs of calibrated sources found in processed epoch images\\
Proc  & PSF     & 480 &  $\sim2$\,days & FITS & PSFs of individual images\\
Merged & Catalog & 24 & $\infty$ & FITS & Catalogs of calibrated sources matched in epoch images\\
Merged & Matrices & 24 & $\infty$ & HDF5 & Calibrated sources matched in epoch images (source vs.~epoch matrices)\\
Coadd & Image & 24 & $\infty$ & FITS & Coadd images of visit \\
Coadd & Mask & 24 & $\infty$ & FITS & Coadd mask images of visit \\
Coadd & Catalog & 24 & $\infty$ & FITS & Catalogs of sources found in coadd images of visit \\
Coadd & PSF & 24 & $\infty$ & FITS & PSFs of coadd images \\
Proc & Asteroids & 1 & $\infty$ & MAT & Asteroids found in visit
\enddata
\tablecomments{Summary of all data products. \#/visit/tel is the number of files 
produced per visit (assuming 20 epochs in a visit) per one telescope. 
Availability indicates the currently estimated lifetime of the product before deletion. 
Individual images can be saved from deletion on demand.
Note that the sub-images and their corresponding catalogs include overlapping regions.
\label{tab:DataProducts}}
\end{deluxetable*}

\subsection{The naming convention}
\label{sec:naming}

All data products follow the same naming conventions,
with the same file name format.
The file names are unique and they
are concatenated from
several strings separated by
underline (``\_'').
The rationale for this convention
is that all file names have the same structure and could be analyzed 
using a single routine that uses simple regular expression commands.
If some sub-string is not relevant to the file, then it appears as an empty string.
In this case, two (or more) successive underlines (e.g., "\_\_") will appear in the file name.
The strings, by their order of appearance in the file names, are:
\begin{itemize}
    \item Project Name --  Project/telescope name. 
    For LAST we use {\tt LAST.$\langle$Node$\rangle$.$\langle$Mount$\rangle$.$\langle$Telescope$\rangle$}. 
    where {\tt Node} is the node index (1 for the first node in Neot-Smadar, Israel). 
    {\tt Mount} for the telescope-mount index (e.g., 1 to 12). {\tt Camera}, for the camera on mount index (e.g., 1 to 4).
    \item Time --  UTC date and time in format {\tt YYYYMMDD.HHMMSS.FFF}.
    \item Filter -- Filter name (e.g., `clear').
    \item Field -- The field ID. For LAST the field ID is a string of the 
    format ddd$\pm$dd, indicating the RA and Dec in decimal degrees.
    \item Counter -- Image counter. For LAST the image counter is usually 1 to 20, 
    indicating the index of the image in the sequence of 20 exposures in a visit.
    \item CCDID -- Detector ID. For LAST this is always 1.
    \item CropID -- Index of the sub-image (1 to 24). 0 or empty string is reserved for the full image.
    \item Type -- One of the following (self-explanatory) image types: {\tt bias}, {\tt dark}, {\tt flat}, {\tt domeflat}, {\tt twflat}, {\tt skyflat}, {\tt fringe}, {\tt focus}, {\tt sci}, {\tt wave}, {\tt test}.
    \item Level -- One of the following strings describing the level of processing: 
    \begin{itemize}
        \item {\tt raw} - A raw image.
        \item {\tt proc} - A single processed image.
        \item {\tt coadd} - The coadd image of the visit or any other sequence of images.
        \item {\tt merged} - Catalogs based on data from multiple epochs.
        \item {\tt ref} - A reference image.
        \item {\tt calib} - A calibration image.
    \end{itemize}
    \item Product -- One of the following keywords describing the file content:
    \begin{itemize}
        \item {\tt Image} - Image data.
        \item {\tt Back} - Background image.
        \item {\tt Var} - Variance image (typically background variance only).
        \item {\tt Mask} - A bit mask image.
        \item {\tt PSF} - A PSF image, or cube.
        \item {\tt Cat} - A Catalog of sources detected in the image.
        \item {\tt TransientsCat} - A Catalog of transient candidates.
        \item {\tt MergedMat} - A merged matrices of sources detected in multiple epochs of the same field.
        \item {\tt Asteroid} - A file containing information on asteroids detected in the image.
        \item {\tt Evt} - A photon event file.
        \item {\tt Spec} - A spectrum.
    \end{itemize}
    \item Version - Processing version.
\end{itemize}

File type extensions are typically, `fits', `hdf5', or `mat'.

\subsection{Individual images}
\label{sec:DP_IndivImages}

The individual image data products include the image itself,
the corresponding bit mask image, the PSF, and a catalog
(Product: `Image', `Mask', 'PSF', and `Cat', respectively).
The images are saved in FITS format, while the catalogs are in FITS binary table format.
The image headers contain additional information, like the limiting magnitude, 
the sky brightness, the fitted photometric zero points, the astrometric World Coordinate System (WCS), and parameters regarding the weather, enclosure, and mount (see Table~\ref{tab:HeaderKeys}).
A single visit of a single telescope yields 1920 files
(20 instances of 24 sub-images, each with 4 data products: [`Image', `Mask', `PSF`, `Cat']).

\subsection{Visit matched sources}
\label{sec:DP_MatchedCat}

The sources in all the catalogs of a single sub-image in a visit are matched and two data products are generated:
(i) A FITS binary table containing the matched sources and some properties including the proper motion fits and variability indicators (see Table~\ref{tab:MergedCat} for columns description);
(ii) A matched source matrices file, including an epoch vs.~source matrix for selected properties like magnitudes and positions (see Table~\ref{tab:MergedCatMat} for matrices description).
This data product is saved in an HDF5 file with multiple datasets.
Each dataset corresponds to a different property (e.g., 'RA', 'MAG\_APER\_3'),
and each such dataset contains a matrix (epoch vs. source).
The magnitudes in these matrices are the calibrated magnitudes but after a relative photometry step.
A single visit of a single telescope yields 48 files (24 sub-images, with 2 data products).

\subsection{Asteroids}
\label{sec:DP_Asteroids}

Data on asteroid candidates are saved with their metadata: 
linear motion fits, source matching, and image cutouts are stored
in {\tt MATLAB} ({\tt mat}) files.
For each field (all sub-images),
a single {\tt mat} file is saved,
where the sub-strings in the file name
are set to:
Type=`science',
Level=`proc',
Product=`Asteroids'.


\subsection{Coadd images}
\label{sec:DP_Coadd}

The images and the bit mask of each visit are coadded.
The main products of this step
are the coadded image, the coadded bit mask, the catalog of the coadded image,
and a PSF of the coadded image.
These data products are similar to those generated for individual images.
A minor difference is that the source-dominated noise bit is populated only for the coadded images.
This step ends up with 96 files per visit per telescope (24 sub-images, with four data products).

\section{On Sky Performances}
\label{sec:Performances}

Here we present some initial results obtained using the LAST pipeline
on the data collected by the LAST system and LAST prototype telescopes.
These results mainly demonstrate the pipeline performances, 
while the camera properties and site-related parameters are discussed 
in \cite{Ofek+2023PASP_LAST_Overview}.

\subsection{Astrometric precision}
\label{sec:Perf_astrometry}

\begin{figure}
\includegraphics[width=8cm]{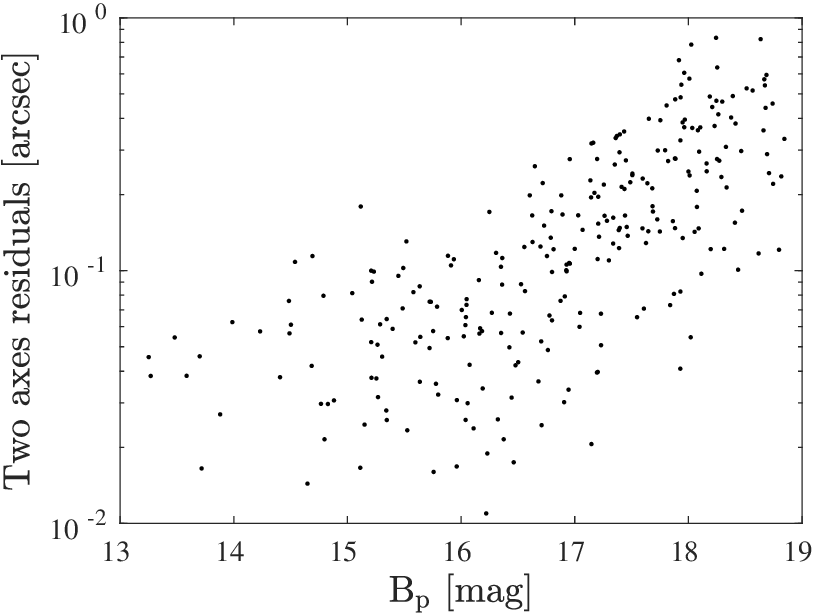}
\caption{Astrometric two axes rms vs.~$B_{\rm p}$ magnitude for a random single 20\,s exposure LAST image.
\label{fig:L289_SubIm5_Astrometry_rmsMag}}
\end{figure}

\begin{figure}
\includegraphics[width=8cm]{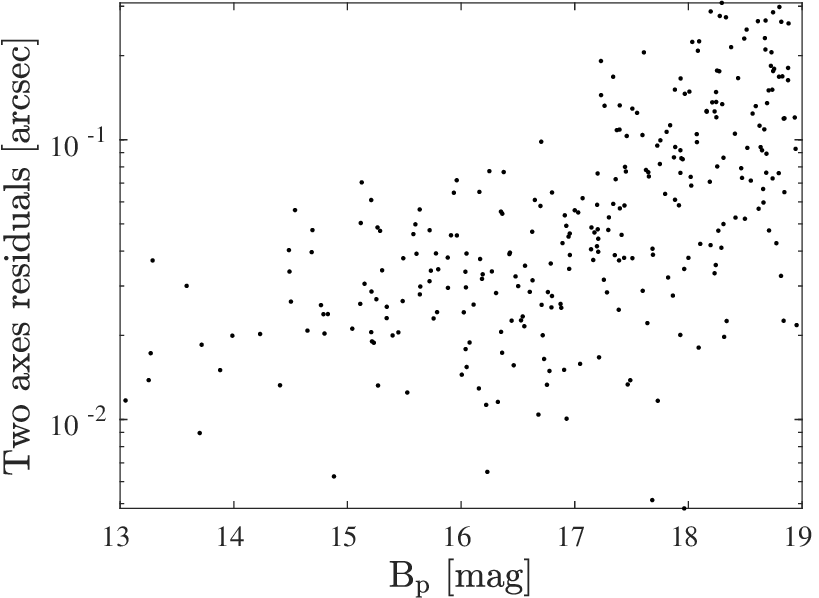}
\caption{Like Figure~\ref{fig:L289_SubIm5_Astrometry_rmsMag}, but for a coadd image of $20\times20$\,s exposures.
\label{fig:L289309_SubIm5_Astrometry_rmsMag}}
\end{figure}

In Figure~\ref{fig:L289_SubIm5_Astrometry_rmsMag} 
we show the two-axes astrometric residuals, compared to GAIA-DR3, as a function of star magnitude
for a single LAST image, while in Figure~\ref{fig:L289309_SubIm5_Astrometry_rmsMag} 
we show the same for a coadd image.
Following \cite{Ofek2019_Astrometry_Code} we define the asymptotic rms by fitting
a 3rd order polynomial to the rms vs.~magnitude distribution, and choose
the minimum of the fitted function that is found between the saturation limit
and the faintest sources.

\begin{figure}
\includegraphics[width=8cm]{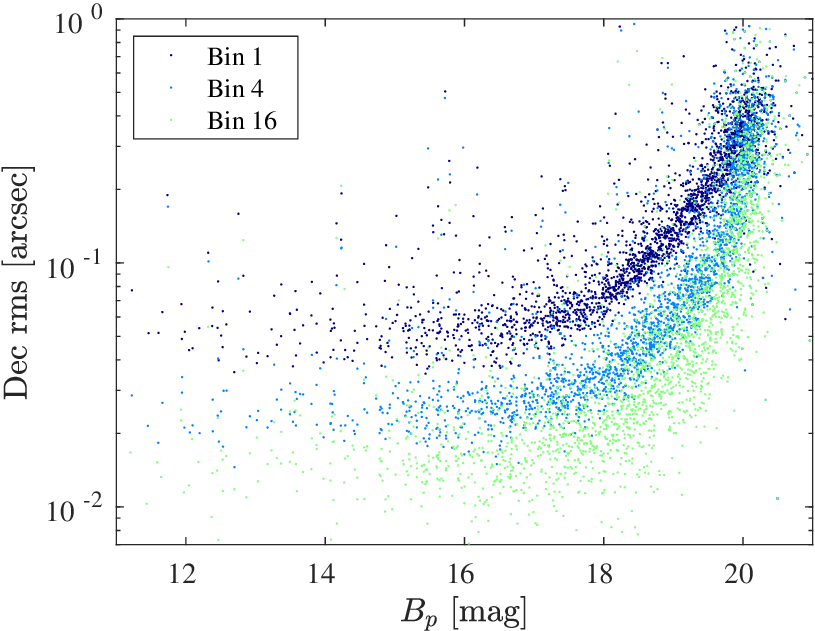}
\caption{The one-axis Declination rms of the relative astrometry solution,
measured over 100 epochs of 20\,s exposures, vs.~the GAIA $B_{\rm p}$ magnitude.
The color represents single epochs (dark blue; Bin1),
an average over 4 epochs (blue; Bin4),
and an average over 16 epochs (green; Bin16).
\label{fig:RelAstrometry_Bin1_4_16}}
\end{figure}

The improvement in astrometric accuracy between a 20\,s
image and a coaddition of 20, 20\,s, images is only a factor of about two,
instead of $\sqrt{20}$.
This likely indicates that our image registration process adds astrometric noise,
and we are investigating this for future improvement.
However, more precise astrometry is possible by averaging
astrometric measurements obtained in different images.
Figure~\ref{fig:RelAstrometry_Bin1_4_16} shows
the relative astrometry rms in declination
as a function of $B_{\rm p}$ magnitude,
calculated also by averaging the astrometry of individual images.
Dots with different colors represent various binning of successive images.
Dark blue (Bin1) is for no binning -- i.e., precision obtained on a 20\,s time scale;
Blue (Bin4) is for binning of four points (80\,s time scale);
and green (Bin16) is for binning of 16 points (320\,s time scale).
At least up to a binning of 16 epochs,
the relative astrometry seems to improve roughly 
like the square root of the number of epochs.

\subsection{Photometric calibration}
\label{sec:Perf_photCalib}

\begin{figure}
\includegraphics[width=8cm]{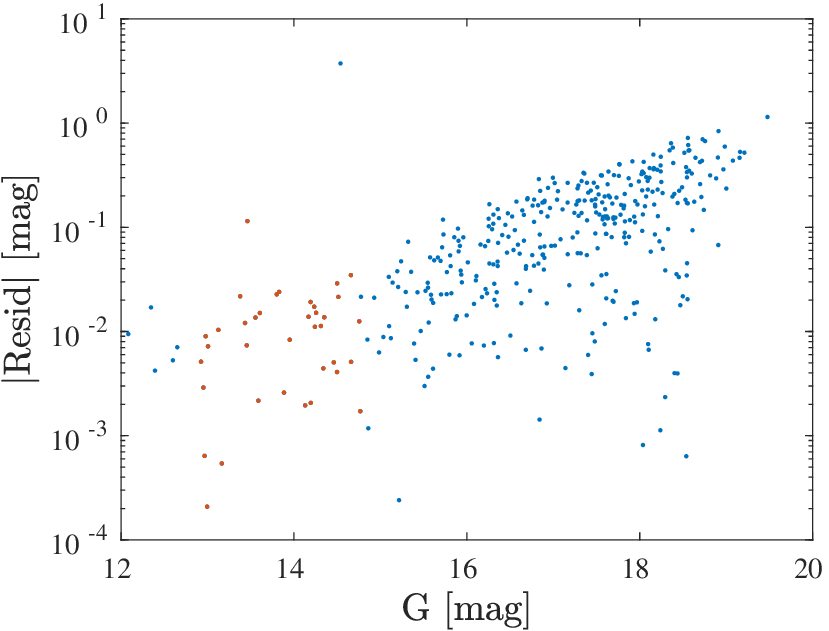}
\caption{The photometric residuals, of the absolute photometric calibration, compared to GAIA-DR3 $B_{\rm p}$ magnitudes,
as a function of magnitude.
Shown for a single exposure sub-image of a random field.
The orange dots are for sources with $S/N$ in the range of about 100 to 1000 that were used
for the calibration. The typical residuals at the bright end are of the order of 1.5\%.
\label{fig:SingleImagePhotCalib_rms_vs_mag}}
\end{figure}

The LAST data is calibrated against the GAIA catalog (\S\ref{sec:PhotCalib}), 
and all the magnitudes are in the native LAST AB system 
(\citealt{Oke+Gunn1983ApJ_FluxCalib_ABmagSystem}; see \S\ref{sec:PhotCalib}).
The current LAST pipeline uses a simple calibration method (\S\ref{sec:PhotCalib}),
while a more sophisticated algorithm is under development.
In Figure~\ref{fig:SingleImagePhotCalib_rms_vs_mag} we present 
the absolute value of the residuals between the fitted and measured magnitudes, 
for a single LAST exposure, as a function of magnitude.
The typical calibration is good to about 1.5\% at the bright end.

\subsection{Relative photometry and variability}
\label{sec:perf_relPhotVar}

\begin{figure}
\includegraphics[width=8cm]{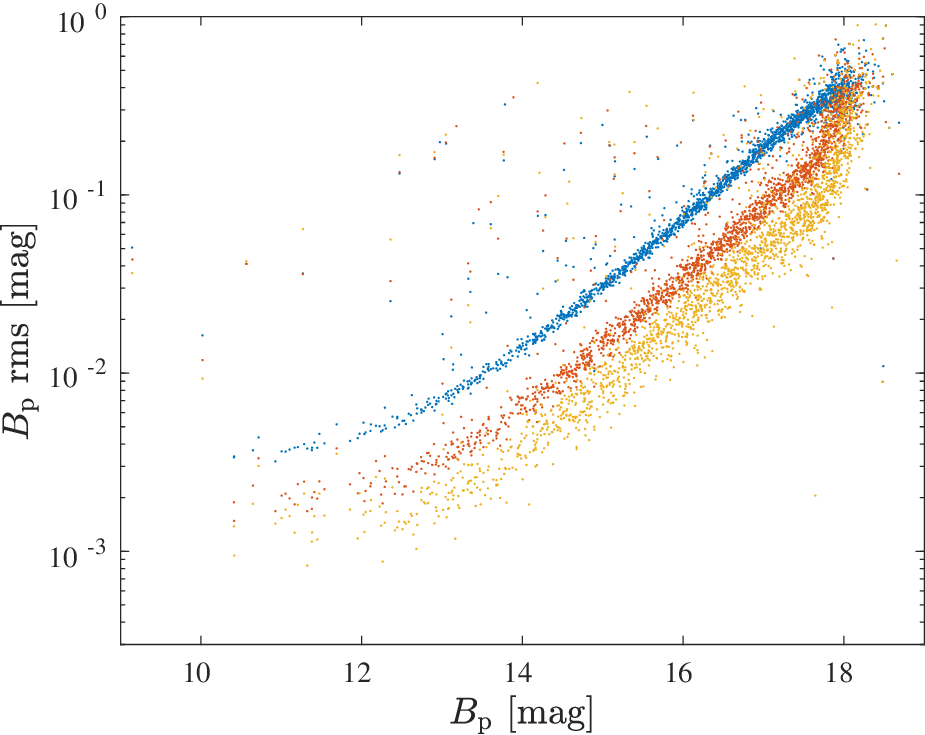}
\caption{The rms of the relative photometry solution, 
measured over 200 epochs, vs.~the GAIA $B_{\rm p}$ magnitude.
The color represents single epochs (blue; Bin1),
an average over 4 epochs (orange; Bin4),
and an average over 16 epochs (yellow; Bin16).
The photometry is based on the 6-pixel radius aperture photometry.
Therefore, its performances are sub-optimal at the faint end.
\label{fig:L289309_SubIm5_RelPhot_rmsMag}}
\end{figure}

The visit successive exposures of the same field are
calibrated using a relative photometry technique (\S\ref{sec:MergedCat}).
In Figure~\ref{fig:L289309_SubIm5_RelPhot_rmsMag}
we present an example for the rms vs.~GAIA $B_{\rm p}$ magnitude
derived from relative photometry calibration and  measured over 200 successive images.
This is done using aperture photometry with a 6-pixel radius, 
and hence its performance on faint objects is not optimal.
Dots of different colors represent various binning of successive images.
Blue (Bin1) is for no binning -- i.e., the precision obtained on a 20\,s time scale is shown;
Orange (Bin4) is for binning of four points (80\,s time scale);
and yellow (Bin16) is for binning of 16 points (320\,s time scale).
On time-scales of 60-min, we consistently achieve a precision of 3--4\,milimag per 20\,s exposure.
Averaging over larger time scales (i.e., 80\,s, 320\,s) improved the precision.
At least up to binning of 16 epochs,
the relative photometry seems to improve roughly like the square root of the number of epochs.
The lines agree well with the theoretical Poisson noise plus about $\sim1\,$\,millimag of noise (in 20\,s exposures).
The extra noise is likely dominated by scintillation noise (e.g., \citealt{Young1967AJ_Photometry_Scintialtions_ReigerTheoryConfirmation}; \citealt{Osborn+2015MNRAS_Photometry_Scintilation}), and flat field errors.

\subsection{Coaddition image depth}
\label{sec:perf_coadd}

The limiting magnitude of the coadd images is stored in the coadd image header.
From several nights of data we verified that, as expected (\citealt{Ofek+2023PASP_LAST_Overview}), 
the 20 exposure coadd image depth is about 1.4\,magnitude deeper
than of the individual exposures.
The difference from the expected improvement of $2.5\log_{10}{\sqrt{20}}\cong 1.6$\,mag, is due to readout noise.

\section{Timing}
\label{sec:Timing}

An important requirement for the LAST pipeline is its efficiency.
The main reason for this is that in order to simplify the observatory design
we would like to use limited computational resources.
Specifically, a LAST node uses 24 high-spec desktop computers, each with 30 cores and 256\,GB RAM.
Therefore, we put efforts into the optimization of the code and minimization of its run time.
Performance improvements, relative to, e.g., the original code presented in \cite{Ofek2014_MAAT}, 
have been achieved in many aspects of the code,
including astrometry, PSF fitting, source matching, and background estimation.
For example, the source finding and measuring routine is about 30 times faster than
{\tt SExtractor} (\citealt{Bertin+1996_SExtractor}), 
when running on the same images with the same computational resources.
Another example is that the FITS writer function's run time on test images
is about three times shorter compared to an implementation using 
{\tt CFITSIO} (\citealt{Pence+2010ascl.soft_CFITSIO}).

The resources required for running the second-step pipeline are considerably less
than those required for the first step pipeline, described in this paper.
Therefore, from a design point of view, they do not pose a big challenge in terms of computation time.

\begin{figure*}
\includegraphics[width=18cm]{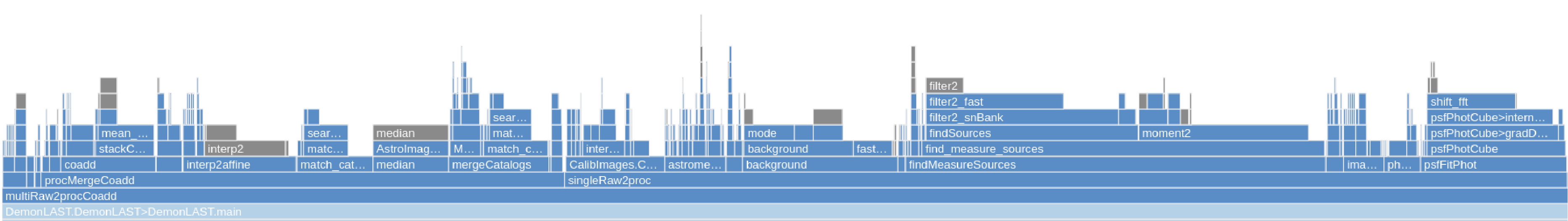}
\caption{A profiler flame plot for the pipeline running on a single visit, with 20 exposures, from a single LAST telescope. Data-saving timing is excluded.
\label{fig:Profiler}}
\end{figure*}

Figure~\ref{fig:Profiler} shows a profiling flame graph for processing a single visit
from a single telescope. The computational load is dominated by saving of the data 
products (not shown in the Figure), and by arithmetic calculations on large matrices, while the overhead from 
the object-oriented design and data management is small.

A few examples: processing of the 480 individual sub-images in a visit, 
of a high-Galactic-latitude field, takes 50\% of the run time,
while working on the merged catalogs and coadd images take about 30\% of the run time,
and 8\% of the run time is spent on writing the data products.
For a high-Galactic-latitude and a low-density field, 
the run time of the pipeline on a single visit is less than 5\,min.
However, for very high-density fields the run time can be significantly longer.
Therefore, we are putting efforts into making the code even more efficient.

\section{Conclusion}
\label{sec:Conclusion}

We presented the Large Array Survey Telescope pipeline-I, which is responsible for processing
a single LAST visit -- e.g., $N$ exposures of a single visit of a single field.

The LAST observational strategy provides an opportunity to study the short-term variability of celestial objects, 
and to screen a large fraction of asteroids and satellite glints in a single visit.
Our pipeline takes advantage of this strategy to yield data products that will 
allow us to explore short-time scale variability and motion of the objects of interest.

A challenge related to any pipeline that handles a high data rate, 
is that the number and size of data products are overwhelming.
For example, a single LAST telescope produces about 2000 data product files for each visit (400\,s).
In order to avoid excessive read times and storage costs
for data files, we avoid using ASCII tables, and our data is kept exclusively in binary files.
Furthermore, given that opening, reading, and closing data files 
is an expensive operation, it is desirable to store 
information in a diversified manner, and store smaller data files with specific products.

We plan to release the LAST data products periodically. 
These data releases will include processed images and catalogs described in this paper 
and in the forthcoming paper~II.
Since the LAST pipeline is constantly updated, we are keeping a live version of this document\footnote{{\url https://www.overleaf.com/read/dymzqrphqqfy
}},
as well as online wiki pages describing the tools used by this pipeline\footnote{{\url https://github.com/EranOfek/AstroPack/wiki}}.

E.O.O. is grateful for the support of
grants from the 
Willner Family Leadership Institute,
André Deloro Institute,
Paul and Tina Gardner,
The Norman E Alexander Family M Foundation ULTRASAT Data Center Fund,
Israel Science Foundation,
Israeli Ministry of Science,
Minerva,
BSF, BSF-transformative, NSF-BSF,
Israel Council for Higher Education (VATAT),
Sagol Weizmann-MIT,
Yeda-Sela, and the
Rosa and Emilio Segre Research Award.
V.F.R. acknowledges support from the German Science Foundation DFG, via the Collaborative Research Center
SFB1491: Cosmic Interacting Matters - from Source to Signal.

\bibliography{papers.bib}
\bibliographystyle{aasjournal}

\end{document}